\documentclass[journal,twocolumn]{IEEEtran}
\usepackage{epsfig,graphics,color,subfigure,wrapfig,hyperref,url}
\usepackage{amssymb,amsmath,algorithm,times}
\usepackage{slashbox}

\begin{document}
\title{Modified Basis Pursuit Denoising(Modified-BPDN) for noisy compressive sensing with partially known support}
\author{Wei Lu and Namrata Vaswani \\Department of Electrical and Computer Engineering, Iowa State University, Ames, IA \\ \{luwei,namrata\}@iastate.edu}

\maketitle

\newcommand{\Dnum}{D_{num}}
\newcommand{\pss}{p^{**,i}}
\newcommand{\fr}{f_{r}^i}

\newcommand{\A}{{\cal A}}
\newcommand{\Z}{{\cal Z}}
\newcommand{\B}{{\cal B}}
\newcommand{\R}{{\cal R}}
\newcommand{\reg}{{\cal G}}
\newcommand{\const}{\mbox{const}}

\newcommand{\trace}{\mbox{trace}}

\newcommand{\hsim}{{\hspace{0.0cm} \sim  \hspace{0.0cm}}}
\newcommand{\he}{{\hspace{0.0cm} =  \hspace{0.0cm}}}

\newcommand{\vect}[2]{\left[\begin{array}{cccccc}
     #1 \\
     #2
   \end{array}
  \right]
  }

\newcommand{\matr}[2]{ \left[\begin{array}{cc}
     #1 \\
     #2
   \end{array}
  \right]
  }
\newcommand{\vc}[2]{\left[\begin{array}{c}
     #1 \\
     #2
   \end{array}
  \right]
  }

\newcommand{\gdot}{\dot{g}}
\newcommand{\Cdot}{\dot{C}}
\newcommand{\re}{\mathbb{R}}
\newcommand{\n}{{\cal N}}  
\newcommand{\N}{{\overrightarrow{\bf N}}}  
\newcommand{\chat}{\tilde{C}_t}
\newcommand{\chati}{\chat^i}

\newcommand{\cmin}{C^*_{min}}
\newcommand{\twi}{\tilde{w}_t^{(i)}}
\newcommand{\twj}{\tilde{w}_t^{(j)}}
\newcommand{\wi}{{w}_t^{(i)}}
\newcommand{\twio}{\tilde{w}_{t-1}^{(i)}}

\newcommand{\tWi}{\tilde{W}_n^{(m)}}
\newcommand{\tWj}{\tilde{W}_n^{(k)}}
\newcommand{\Wi}{{W}_n^{(m)}}
\newcommand{\tWio}{\tilde{W}_{n-1}^{(m)}}

\newcommand{\ds}{\displaystyle}

\newcommand{\SAR}{S$\!$A$\!$R }
\newcommand{\MAR}{MAR}
\newcommand{\MMRF}{MMRF}
\newcommand{\AR}{A$\!$R }
\newcommand{\GMRF}{G$\!$M$\!$R$\!$F }
\newcommand{\DTM}{D$\!$T$\!$M }
\newcommand{\MSE}{M$\!$S$\!$E }
\newcommand{\RCS}{R$\!$C$\!$S }
\newcommand{\uomega}{\underline{\omega}}
\newcommand{\y}{v}
\newcommand{\x}{w}
\newcommand{\lu}{\mu}
\newcommand{\g}{g}
\newcommand{\s}{{\bf s}}
\newcommand{\bft}{{\bf t}}
\newcommand{\refmap}{{\cal R}}
\newcommand{\totrefl}{{\cal E}}
\newcommand{\beq}{\begin{equation}}
\newcommand{\eeq}{\end{equation}}
\newcommand{\bdm}{\begin{displaymath}}
\newcommand{\edm}{\end{displaymath}}
\newcommand{\hatz}{\hat{z}}
\newcommand{\hatu}{\hat{u}}
\newcommand{\tilz}{\tilde{z}}
\newcommand{\tilu}{\tilde{u}}
\newcommand{\hhatz}{\hat{\hat{z}}}
\newcommand{\hhatu}{\hat{\hat{u}}}
\newcommand{\tilc}{\tilde{C}}
\newcommand{\hatc}{\hat{C}}
\newcommand{\tim}{n}

\newcommand{\ssp}{\renewcommand{\baselinestretch}{1.0}}
\newcommand{\defd}{\mbox{$\stackrel{\mbox{$\triangle$}}{=}$}}
\newcommand{\goes}{\rightarrow}
\newcommand{\tends}{\rightarrow}
\newcommand{\defn}{\triangleq} 
\newcommand{\se}{&=&}
\newcommand{\sdefn}{& \defn  &}
\newcommand{\sle}{& \le &}
\newcommand{\sge}{& \ge &}
\newcommand{\plusminus}{\stackrel{+}{-}}
\newcommand{\Ey}{E_{Y_{1:t}}}
\newcommand{\ey}{E_{Y_{1:t}}}

\newcommand{\equivto}{\mbox{~~~which is equivalent to~~~}}
\newcommand{\nonzero}{i:\pi^n(x^{(i)})>0}
\newcommand{\nonzeroc}{i:c(x^{(i)})>0}

\newcommand{\supn}{\sup_{\phi:||\phi||_\infty \le 1}}
\newtheorem{theorem}{Theorem}
\newtheorem{lemma}{Lemma}
\newtheorem{corollary}{Corollary}
\newtheorem{definition}{Definition}
\newtheorem{remark}{Remark}
\newtheorem{example}{Example}
\newtheorem{ass}{Assumption}
\newtheorem{fact}{Fact}
\newtheorem{heuristic}{Heuristic}
\newcommand{\eps}{\epsilon}
\newcommand{\bd}{\begin{definition}}
\newcommand{\ed}{\end{definition}}
\newcommand{\udq}{\underline{D_Q}}
\newcommand{\td}{\tilde{D}}
\newcommand{\epsinv}{\epsilon_{inv}}
\newcommand{\al}{\mathcal{A}}

\newcommand{\bfx} {\bf X}
\newcommand{\bfy} {\bf Y}
\newcommand{\bfz} {\bf Z}
\newcommand{\ddas}{\mbox{${d_1}^2({\bf X})$}}
\newcommand{\ddbs}{\mbox{${d_2}^2({\bfx})$}}
\newcommand{\dda}{\mbox{$d_1(\bfx)$}}
\newcommand{\ddb}{\mbox{$d_2(\bfx)$}}
\newcommand{\xinc}{{\bfx} \in \mbox{$C_1$}}
\newcommand{\eqa}{\stackrel{(a)}{=}}
\newcommand{\eqb}{\stackrel{(b)}{=}}
\newcommand{\eqe}{\stackrel{(e)}{=}}
\newcommand{\leqc}{\stackrel{(c)}{\le}}
\newcommand{\leqd}{\stackrel{(d)}{\le}}

\newcommand{\leqa}{\stackrel{(a)}{\le}}
\newcommand{\leqb}{\stackrel{(b)}{\le}}
\newcommand{\leqe}{\stackrel{(e)}{\le}}
\newcommand{\leqf}{\stackrel{(f)}{\le}}
\newcommand{\leqg}{\stackrel{(g)}{\le}}
\newcommand{\leqh}{\stackrel{(h)}{\le}}
\newcommand{\leqi}{\stackrel{(i)}{\le}}
\newcommand{\leqj}{\stackrel{(j)}{\le}}

\newcommand{\w}{{W^{LDA}}}
\newcommand{\halpha}{\hat{\alpha}}
\newcommand{\hsigma}{\hat{\sigma}}
\newcommand{\slmax}{\sqrt{\lambda_{max}}}
\newcommand{\slmin}{\sqrt{\lambda_{min}}}
\newcommand{\lmax}{\lambda_{max}}
\newcommand{\lmin}{\lambda_{min}}

\newcommand{\da} {\frac{\alpha}{\sigma}}
\newcommand{\chka} {\frac{\check{\alpha}}{\check{\sigma}}}
\newcommand{\sumo}{\sum _{\underline{\omega} \in \Omega}}
\newcommand{\distance}{d\{(\hatz _x, \hatz _y),(\tilz _x, \tilz _y)\}}
\newcommand{\col}{{\rm col}}
\newcommand{\rcs}{\sigma_0}
\newcommand{\CalR}{{\cal R}}
\newcommand{\df}{{\delta p}}
\newcommand{\dq}{{\delta q}}
\newcommand{\dZ}{{\delta Z}}
\newcommand{\pprime}{{\prime\prime}}

\newcommand{\vn}{N}

\newcommand{\bv}{\begin{vugraph}}
\newcommand{\ev}{\end{vugraph}}
\newcommand{\bi}{\begin{itemize}}
\newcommand{\ei}{\end{itemize}}
\newcommand{\ben}{\begin{enumerate}}
\newcommand{\een}{\end{enumerate}}
\newcommand{\be}{\protect\[}
\newcommand{\ee}{\protect\]}
\newcommand{\bean}{\begin{eqnarray*} }
\newcommand{\eean}{\end{eqnarray*} }
\newcommand{\bea}{\begin{eqnarray} }
\newcommand{\eea}{\end{eqnarray} }
\newcommand{\nn}{\nonumber}
\newcommand{\ba}{\begin{array} }
\newcommand{\ea}{\end{array} }
\newcommand{\ep}{\mbox{\boldmath $\epsilon$}}
\newcommand{\epp}{\mbox{\boldmath $\epsilon '$}}
\newcommand{\Lep}{\mbox{\LARGE $\epsilon_2$}}
\newcommand{\und}{\underline}
\newcommand{\pdif}[2]{\frac{\partial #1}{\partial #2}}
\newcommand{\odif}[2]{\frac{d #1}{d #2}}
\newcommand{\dt}[1]{\pdif{#1}{t}}
\newcommand{\urho}{\underline{\rho}}

\newcommand{\spc}{{\cal S}}
\newcommand{\tspc}{{\cal TS}}

\newcommand{\uv}{\underline{v}}
\newcommand{\us}{\underline{s}}
\newcommand{\uc}{\underline{c}}
\newcommand{\utheta}{\underline{\theta}^*}
\newcommand{\ualpha}{\underline{\alpha^*}}

\newcommand{\uxy}{\underline{x}^*}
\newcommand{\uxyj}{[x^{*}_j,y^{*}_j]}
\newcommand{\arcl}[1]{arclen(#1)}
\newcommand{\one}{{\mathbf{1}}}

\newcommand{\uxyjt}{\uxy_{j,t}}
\newcommand{\E}{\mathbb{E}}

\newcommand{\rhomat}{\left[\begin{array}{c}
                        \rho_3 \ \rho_4 \\
                        \rho_5 \ \rho_6
                        \end{array}
                   \right]}
\newcommand{\deltat}{\tau} 
\newcommand{\deltatt}{\Delta t_1}
\newcommand{\ceil}[1]{\ulcorner #1 \urcorner}

\newcommand{\xxi}{x^{(i)}}
\newcommand{\txi}{\tilde{x}^{(i)}}
\newcommand{\txj}{\tilde{x}^{(j)}}

\newcommand{\mi}[1]{{#1}^{(m,i)}}

\begin{abstract}
In this work, we study the problem of reconstructing a sparse signal from a limited number of linear `incoherent' noisy measurements, when a part of its support is known. The known part of the support may be available from prior knowledge or from the previous time instant (in applications requiring recursive reconstruction of a time sequence of sparse signals, e.g. dynamic MRI). We study a modification of Basis Pursuit Denoising (BPDN) and bound its reconstruction error. A key feature of our work is that the bounds that we obtain are computable. Hence, we are able to use Monte Carlo to study their average behavior as the size of the unknown support increases. We also demonstrate that when the unknown support size is small,  modified-BPDN bounds are much tighter than those for BPDN,  and hold under much weaker sufficient conditions (require fewer measurements).
\end{abstract}

\section{Introduction}
In this work, we study the problem of reconstructing a sparse signal from a limited number of linear `incoherent' noisy measurements, {\em when a part of its support is known}. In practical applications, this may be obtained from prior knowledge, e.g. it can be the lowest subband of wavelet coefficients for medical images which are sparse in the wavelet basis. Alternatively when reconstructing time sequences of sparse signals, e.g. in a real-time dynamic MRI application, it could be the support estimate from the previous time instant.

In \cite{modcs}, we introduced modified-CS for the noiseless measurements' case. Sufficient conditions for exact reconstruction were derived and it was argued that these are much weaker than those needed for CS. Modified-CS-residual, which combines the modified-CS idea with CS on LS residual (LS-CS) \cite{analyzelscs}, was introduced for noisy measurements in \cite{modcsMRI} for a real-time dynamic MRI reconstruction application. In this paper, we bound the recosntruction error of a simpler special case of modified-CS-residual, which we call modified-BPDN. We use a strategy similar to the results of \cite{justrelax} to bound the reconstruction error and hence, just like in \cite{justrelax},  {\em the bounds we obtain are computable}. We are thus able to use Monte Carlo to study the average behavior of the reconstruction error bound as the size of the unknown support, $\Delta$, increases or as the size of the support itself, $N$, increases. We also demonstrate that modified-BPDN bounds are much smaller than those for BPDN (which corresponds to $|\Delta|=|N|$) and hold under much weaker sufficient conditions (require fewer measurements).

In parallel and independent work recently posted on Arxiv, \cite{iBPDN} also proposed an approach related to modified-BPDN and bounded its error. Their bounds are based on Candes' results and hence are not computable. Other related work includes \cite{timevarying} (which focusses on the time series case and mostly studies the time-invariant support case) and \cite{weightedl1} (studies the noiseless measurements' case and assumes probabilistic prior knowledge).
\subsection{Problem definition}
We obtain an $n$-length measurement vector $y$ by
\begin{equation}
y=Ax+w
\end{equation}
Our problem is to reconstruct the $m$-length sparse signal $x$ from the measurement $y$ with $m>n$. The measurement is obtained from an $n\times m$ measurement matrix $A$ and corrupted by a $n$-length vector noise $w$. The support of $x$ denoted as $N$ consists of three parts: $N\triangleq T\cup \Delta \setminus \Delta_e$ where  $\Delta$ and $T$ are disjoint and $\Delta_e \subseteq T$. $T$ is the known part of support while $\Delta_e$ is the error in the known part of support and $\Delta$ is the unknown part. We also define $N_e \triangleq T \cup \Delta=N\cup \Delta_e$.

\textit{Notation:} We use $'$ for conjugate transpose. For any set $T$ and
vector $b$, we have $(b)_T$ to denote a sub-vector
containing the elements of $b$ with indices in $T$.
$\|b\|_k$ means the $l_k$ norm of the vector $b$. $T^c$ denotes the complement of set $T$ and $\emptyset$ is the empty set. For the
matrix $A$, $A_T$ denotes the sub-matrix by extracting columns of
$A$ with indices in $T$. The matrix norm $\|A\|_{p}$, is
defined as
\begin{equation}
\|A\|_{p}\triangleq \max_{x\neq 0}\frac{\|Ax\|_p}{\|x\|_p}\nonumber
\end{equation}

We also define $\delta_{S}$ to be the $S$-restricted isometry constant and $\theta_{S,S'}$ to be the $S,S'$ restricted orthogonality constant as in \cite{decodinglp}.
\section{Bounding modified-BPDN}
In this section, we introduce modified-BPDN and derive the bound for its reconstruction error.
\subsection{Modified-BPDN}
In \cite{modcs}, equation (5) gives the modified-CS algorithm under noiseless measurements. We relax the equality constraint of this equation to propose modified-BPDN algorithm using a modification of the BPDN idea\cite{BPDN}. We solve
\begin{equation}
\min_b \ \frac{1}{2}\|y-Ab\|_2^2+\gamma\|b_{T^c}\|_1 \label{modcsnoisy}
\end{equation}
Then the solution to this convex optimization problem $\hat{x}$ will be the reconstructed signal of the problem. In the following two subsections, we bound the reconstruction error.
\subsection{Bound of reconstruction error}
We now bound the reconstruction error. We use a strategy similar to \cite{justrelax}. We define the function
\begin{equation}
\ \ L(b)=\frac{1}{2}\|y-Ab\|_2^2+\gamma\|b_{T^c}\|_1 \label{modcsnoisyfunction}
\end{equation}
Look at the solution of the problem (\ref{modcsnoisy}) over all vectors supported on $N_e$. If $A_{N_e}$ has full column rank, the function $L(b)$ is strictly convex when minimizing it over all $b$ supported on $N_e$ and then it will have a unique minimizer. We denote the unique minimizer of function $L(b)$ over all $b$ supported on $N_e$ as
\begin{equation}
\tilde{b}=[\tilde{b}'_{N_e} \ \ \textbf{0}'_{N_e^c}]\label{minimizer}
\end{equation}
Also, we denote the genie-aided least square estimate supported on $N_e$ as
\begin{equation}
c:=[c'_{N_e} \ \ \textbf{0}'_{N_e^c}] \text{   where  }c_{N_e}:=(A'_{N_e}A_{N_e})^{-1}
A'_{N_e}y \label{lssolution}
\end{equation}
Since $\|c-x\|_2 \le \frac{\|w\|}{\sqrt{1-\delta_{|N_e|}}}$ is quite small if noise is small and $\delta_{|N_e|}$ is small, we just give the error bound for $\tilde{b}$ with respect to $c$ in the following lemma and will prove that it is also the global unique minimizer under some sufficient condition.
\begin{lemma}
Suppose that $A_{N_e}$ has full column rank, and let $\tilde{b}$ minimize the function $L(b)$ over all vectors supported on $N_e$. We have the following conclusions:
\begin{enumerate}
\item A necessary and sufficient condition for $\tilde{b}$ to be the unique minimizer is that
\begin{displaymath}
c_{N_e}-\tilde{b}_{N_e}= \left [ \begin{array}{c}
-\gamma (A_T'A_T)^{-1}A'_TA_{\Delta}(A'_{\Delta}MA_{\Delta})^{-1}g_{\Delta}\\
\gamma (A'_{\Delta}MA_{\Delta})^{-1}g_{\Delta})
\end{array} \right ]
\end{displaymath}
where $M\triangleq I-A_T(A'_TA_T)^{-1}A'_T$ and $g\in \partial(\|b_{T^c}\|_1)|_{b=\tilde{b}}$.
$\partial (\|b_{T^c}\|_1)$ is the subgradient set of $\|b_{T^c}\|_1$. Thus, $g_T=0$ and $\|g_{\Delta}\|_{\infty}=1$.
\item Error bound in $l_{\infty}$ norm
\begin{eqnarray}
\|\tilde{b}-c\|_{\infty}\le \gamma
\max(\|(A'_TA_T)^{-1}A'_TA_{\Delta}(A'_{\Delta}MA_{\Delta})^{-1}\|_{\infty}\hspace{-8mm}\nonumber\\
,\|(A'_{\Delta}MA_{\Delta})^{-1}\|_{\infty})\hspace{2mm}\label{lsbound}
\end{eqnarray}
\item Error bound in $l_2$ norm
\begin{eqnarray}
\|\tilde{b}-c\|_2\le \gamma \sqrt{|\Delta|}\cdot \quad \quad \quad \quad \quad \quad \quad \quad \quad \quad \quad \quad \quad \quad \quad \nonumber \\
\sqrt{\|(A_T'A_T)^{-1}A_T'A_{\Delta}(A_{\Delta}'MA_{\Delta}^{-1})\|_{2}^2+\|(A_{\Delta}'MA_{\Delta})^{-1}\|_{2}^2}\nonumber\\
\le \gamma \sqrt{|\Delta|}\sqrt{\frac{\theta_{|T|,|\Delta|}^2}{(1-\delta_{|T|})^2}+1}\cdot \frac{1}{1-\delta_{|\Delta|}-\frac{\theta_{|\Delta|,|T|}^2}{1-\delta_{|T|}}}\nonumber
\end{eqnarray}
\end{enumerate}
\end{lemma}
The proof is given in the Appendix.

Next, we obtain sufficient condition under which $\tilde{b}$ is also the unique global minimizer of $L(b)$.
\begin{lemma}
If the following condition
is satisfied, then the problem
(\ref{modcsnoisy}) has a unique minimizer which is equal to $\tilde{b}$ defined in (\ref{minimizer}).
\begin{equation}
\|A'(y-A_{N_e}c_{N_e})\|_{\infty}< \gamma \big{[}1-\max_{\omega\notin
N_e}\|(A'_{\Delta}MA_{\Delta})^{-1}A'_{\Delta}MA_{\omega}\|_1\big{]}\nonumber
\end{equation}
\end{lemma}
The proof of Lemma 2 is in the appendix.

Combining Lemma 1 and 2 and bounding $\|c-x\|$,we get the following Theorem:
\begin{theorem} If $A_{N_e}$ has full column rank and the
following condition is satisfied
\begin{equation}
\|A'(y-A_{N_e}c_{N_e})\|_{\infty}< \gamma \big{[}1-\max_{\omega \notin
N_e}\|(A'_{\Delta}MA_{\Delta})^{-1}A'_{\Delta}MA_{\omega}\|_1\big{]}\label{directcondition}
\end{equation}
then,
\begin{enumerate}
\item Problem (\ref{modcsnoisy}) has a unique minimizer $\tilde{b}$ and it is supported on $N_e$.
\item The unique minimizer $\tilde{b}$ satisfies
\begin{eqnarray}
\|\tilde{b}-x\|_{\infty}\le\gamma \max( \|(A'_TA_T)^{-1}A'_TA_{\Delta}(A'_{\Delta}MA_{\Delta})^{-1}\|_{\infty} \nonumber\\
,\|(A'_{\Delta}MA_{\Delta})^{-1}\|_{\infty})+\|(A_{N_e}'A_{N_e})^{-1}A_{N_e}'\|_{\infty}\|w\|_{\infty}\label{errorbounddelta}
\end{eqnarray}
and
\begin{align}
&\|\tilde{b}-x\|_2\le \|(A_{N_e}'A_{N_e})^{-1}A_{N_e}'\|_{2}\|w||_{2}+\gamma \sqrt{|\Delta|}\cdot  \nonumber \\ &\sqrt{\|(A_T'A_T)^{-1}A_T'A_{\Delta}(A_{\Delta}'MA_{\Delta}^{-1})\|_{2}^2+\|(A_{\Delta}'MA_{\Delta})^{-1}\|_{2}^2}\label{l2bound}\\
&\le  \gamma \sqrt{|\Delta|}\sqrt{\frac{\theta_{|T|,|\Delta|}^2}{(1-\delta_{|T|})^2}+1}\cdot \frac{1}{1-\delta_{|\Delta|}-\frac{\theta_{|\Delta|,|T|}^2}{1-\delta_{|T|}}}+\frac{\|w\|_2}{\sqrt{1-\delta_{|N_e|}}}\label{ripbound}
\end{align}
\end{enumerate}
\end{theorem}
Now consider BPDN. From theorem 8 of \cite{justrelax}(the same thing also follows by setting $T=\emptyset$ in our result), if $A_N$ has full rank and if
\begin{equation}
\|A'(y-A_{N}(A_N'A_N)^{-1}A_N'y)\|_{\infty}<\gamma [1-\max_{\omega \notin N}\|(A_N'A_N)^{-1}A_N'A_{\omega}\|_1]\label{erc}
\end{equation}
then $\tilde{b}_{BPDN}$
\begin{equation}
\|\tilde{b}_{BPDN}-x\|_{\infty}\le \gamma\|(A_{N}'A_{N})^{-1}\|_{\infty}+\|(A_{N}'A_{N})^{-1}A_{N}'\|_{\infty}\|w\|_{\infty}
\end{equation}
Similarly, we can have the $l_2$ norm bound of BPDN is
\begin{equation}
\|\tilde{b}_{BPDN}-x\|_2 \le \gamma \sqrt{|N|} \frac{1}{1-\delta_{|N|}}+\frac{\|w\|_2}{\sqrt{1-\delta_{|N|}}}\label{BPDNl2bound}
\end{equation}
Compare (\ref{ripbound}) and (\ref{BPDNl2bound}) for the case when $|\Delta|=|\Delta_e|=\frac{|N|}{10}$(follows from \cite{modcsMRI}), the second terms are mostly equal. Consider an example assuming that $\delta_{|N|}=0.5$, $\delta_{|\Delta|}=0.1$, $\theta_{|T|,|\Delta|}=0.2$ and $|\Delta|=\frac{1}{10}|N|$ which is practical in real data. Then the bound for BPDN is $2\gamma_{BPDN} \sqrt{|N|}+0.7||w||_2$ and the bound for modified-BPDN approximates to $1.3\gamma_{modBPDN} |\Delta|+0.7||w||_2$. Using a similar argument, $\gamma_{modBPDN}$ which is the smallest $\gamma$ satisfying (\ref{directcondition}), will be smaller than $\gamma_{BPDN}$ which is the smallest $\gamma$ satisfying (\ref{erc}). Since $|\Delta|=\frac{1}{10}|N|$ and $\gamma_{BPDN}$ will be larger than $\gamma_{modBPDN}$, the bound for modified-BPDN will be much smaller than that of BPDN. This is one example, but we do a detailed simulation comparison in the next section using the computable version of the bounds given in (\ref{errorbounddelta}) and (\ref{l2bound}).
\section{Simulation Results}
In this section,
we compare both the computable $l_{\infty}$ and $l_2$ norm bounds for modified-BPDN with those of BPDN
using Monte Carlo simulation. Note that, BPDN is a special case of
modified-BPDN when $\Delta=N$ and $\Delta_e=\emptyset$. Therefore, we do the
following simulation to check the change of error bound when
$|\Delta|$ increases and compare the bounds of modified-BPDN
with those of BPDN.

\begin{figure}[h]
\centering{ \subfigure[\small{$|N|=100,\Delta_e=\emptyset$}]{
\includegraphics [height=4.5cm]{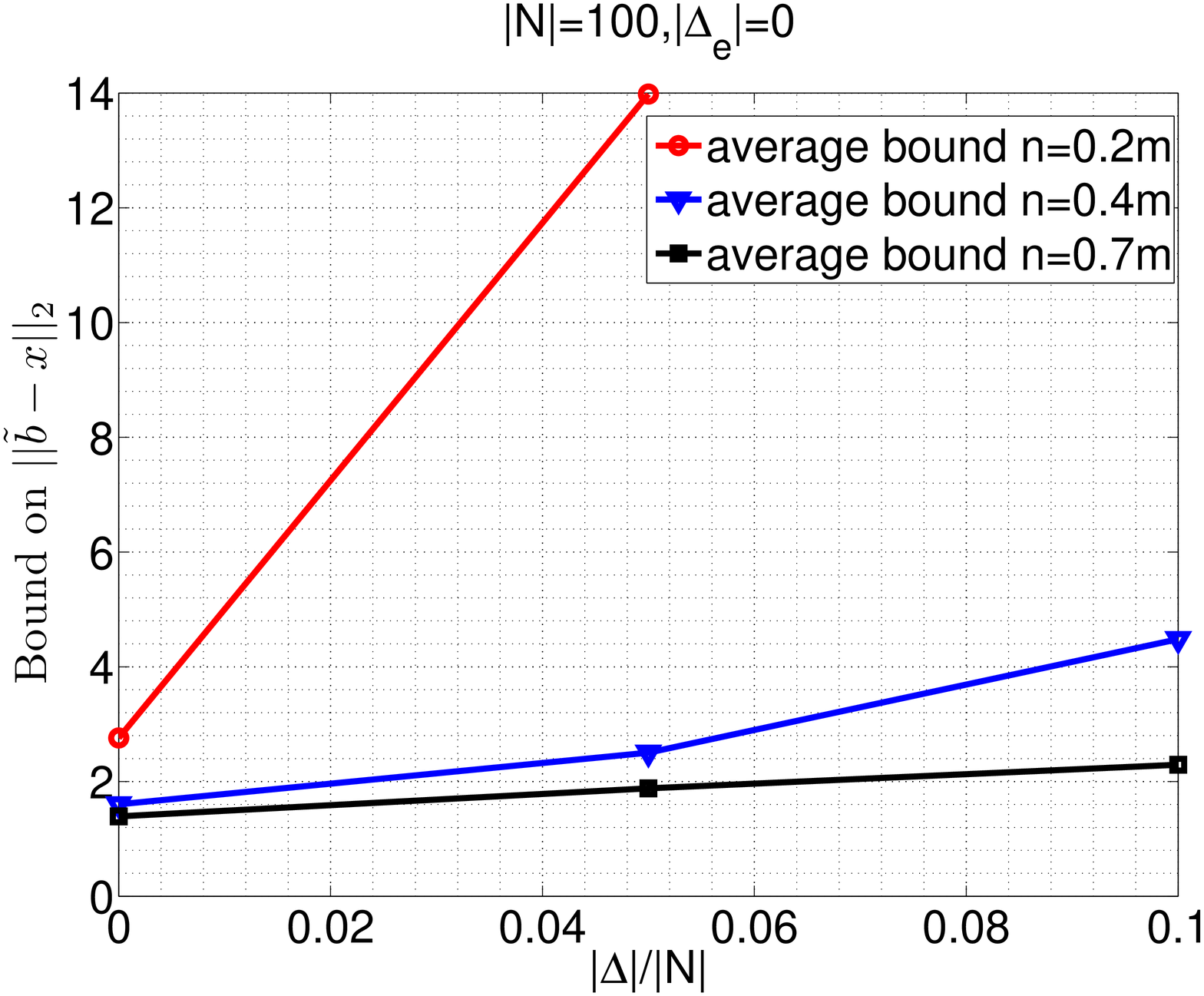}
}
\\
\subfigure[\small{$|N|=100,|\Delta_e|=\frac{1}{10}|N|$}]{
\includegraphics [height=4.5cm]{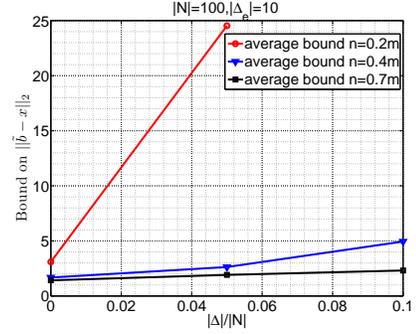}}
}
\caption{\small{The average bound(\ref{l2bound}) on $||\tilde{b}-x||_2$ is plotted. Signal length
$m=1024$ and support size $|N|=100$. For fixed $n$ and $|\Delta_e|$, the bound increases when $|\Delta|$ increases. When number of measurements $n$ increases, the bound decreases. When $n=0.2m$ and $|\Delta|\ge 0.05|N|$, the RHS of (\ref{directcondition}) is negative and thus the bound does not hold.  
We do not plot the case of BPDN($\Delta=N,\Delta_e=\emptyset$) since it requires $n\ge 0.95m$ measurements to make RHS of (\ref{directcondition}) positive.}
} \label{nz100}
\end{figure}

\begin{figure}[h]
\centering{
\includegraphics [height=4.5cm]{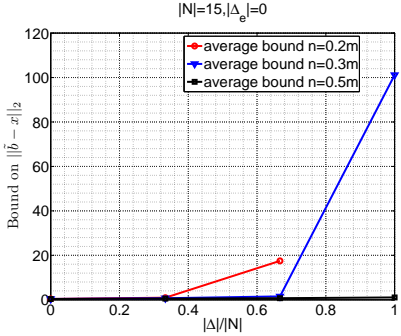}
}
\caption{\small{The average bound(\ref{l2bound}) on $||\tilde{b}-x||_2$ is plotted. Signal length $m=1024$, support size $|N|=15$ and $|\Delta_e|=0$. For fixed $n$, the bound on $||\tilde{b}-x||_2$ increases when $|\Delta|$ increases. When number of measurements $n$ increases, the bound decreases. When $n=0.2m$ and $|\Delta|\ge \frac{2}{3}|N|$, the RHS of (\ref{directcondition}) is negative and thus the bound does not hold.}} \label{nz15}
\end{figure}

We do the simulation as follows:
\begin{enumerate}
\item Fix $m = 1024$ and size of support $|N|$.
\item Select $n$, $|\Delta|$ and $|\Delta_e|$.
\item Generate the $n\times m$ random-Gaussian matrix, $A$ (generate an $n \times m$ matrix with i.i.d. zero mean Gaussian entries and normalize each column to unit $\ell_2$ norm).
\item Repeat the following $\text{tot}=50$ times
\begin{enumerate}
\item  Generate the support, $N$, of size $|N|$, uniformly at random from $[1:m]$.
\item Generate the nonzero elements of the sparse signal $x$ on the support $N$ with i.i.d Gaussian distributed entries with zero mean and variance 100.
Then generate a random i.i.d Gaussian noise $w$ with zero mean and variance $\sigma_w^2$. Compute $y:=Ax+w$.
\item Generate the unknown part of support, $\Delta$, of size $|\Delta|$ uniformly at random from the elements of $N$.
\item Generate the error in known part of support, $\Delta_e$, of size $|\Delta_e|$, uniformly at random from $[1:m] \setminus N$
\item Use $T = N \cup \Delta_e \setminus \Delta$ to compute $\gamma^*$ by
\begin{equation}
 \gamma^*=\frac{\|A'(y-A_{N_e}c_{N_e})\|_{\infty}}{1-\max_{\omega \notin
N_e}\|(A'_{\Delta}MA_{\Delta})^{-1}A'_{\Delta}MA_{\omega}\|_1}\nonumber
 \end{equation}
  and do reconstruction with $\gamma=\gamma^*$ using modified-BPDN to obtain $\hat{x}_{modBPDN}$.
\item Compute the reconstruction error $\|\hat{x}_{modBPDN}-c\|_{\infty}$
\item Compute the $l_{\infty}$ norm bound from (\ref{lsbound}) and the $l_2$ norm bound from (\ref{l2bound}).
\end{enumerate}
\item Compute the average bounds and average error for the given $n$, $|\Delta|$,
$|\Delta_e|$.
\item Repeat for various values of $n$,$|\Delta|$ and $|\Delta_e|$.
\end{enumerate}
Fig.\ref{nz100} shows the average bound(RHS of (\ref{l2bound})) for different $|\Delta|$ when $|N|=100\approx 10\%m$ which is practical for real data as in \cite{modcs,modcsMRI}. The noise variance is $\sigma_w^2=0.001$. We show plots for different choice of $n$. The case $\frac{|\Delta|}{|N|}=1$ in Fig. \ref{nz100} corresponds to BPDN. From the figures, we can observe that when $|\Delta|$ increases, the bounds are increasing. One thing needed to be mentioned is that for BPDN($\Delta=N,\Delta_e=\emptyset$) in this case, the RHS of (\ref{directcondition}) is negative and the bound can only hold when number of measurements $n\ge 0.95m$. Therefore, BPDN is difficult to meet the unique minimizer condition when $|N|$ increases to $0.1m$. However, when $|\Delta|$ is small, modified-BPDN can easily satisfy the condition, even with very few measurements($n=0.2m$
when $|\Delta|=0.05|N|$). 
Hence, the sufficient conditions for modified-BPDN require much fewer measurements than those for
BPDN when $|\Delta|$ is small.
Fig.\ref{nz15} gives another group of results showing average bound(RHS of (\ref{l2bound})) for different
$|\Delta|$ when $|N|=15\approx 1.5\%m$. The noise variance is $\sigma_w^2=0.0003$ and $\Delta_e=\emptyset$. We can also obtain the same conclusions as Fig.\ref{nz100}.  Note that we do not plot the average error and bound for $|\Delta|\ge \frac{2}3|N|$ when $n=0.2m$ since the RHS of (\ref{directcondition}) is negative and thus the bound does not hold. Hence, the more we know the support, the fewer measurements modified-BPDN requires.

In this case, we also compute the average error and the bound (\ref{lsbound}) on $\|\tilde{b}-c\|_{\infty}$. Since $|N_e|=15$ is small and noise is small $\|c-x\|_{\infty}$ will be small and equal for any choice of $|\Delta|$. Thus we just compare $\|\tilde{b}-c\|_{\infty}$ with its upper bound given in (\ref{lsbound}). For the error and bound on $\|\tilde{b}-c\|_{\infty}$, when we fix $n=0.3m$ and $\Delta_e=\emptyset$, the error and the bound are both 0 for $|\Delta|=0$ which verifies that the unique minimizer is equal to the genie-aided least square estimation on support $N$ in this case. For $|\Delta|=\frac{1}{3}|N|$, the error is $0.08$ and the bound is $0.09$. For $|\Delta|=\frac{2}{3}|N|$, the error is $0.21$ and the bound is $0.27$. When $|\Delta|=|N|$ which corresponds to BPDN in this case, the error increases to $3.3$ and the bound increases to $9$. Therefore, we can observe that when $|\Delta|$ increases, both the error and the bound are increasing. Also, we can see the gap between error and bound(gap=bound-error) increases with $|\Delta|$.

From the simulation results, we conclude as follows:
\begin{enumerate}
\item The error and bound increase as $|\Delta|$ increases.
\item The error and bound increase as $|N|$ increases.
\item The gap between the error and bound increases as $|\Delta|$ increases.
\item The error and bound decrease as $n$ increases.
\item For real data, $|N|\approx 0.1m$. In this case, BPDN needs $n\ge 0.95m$ to apply the bound while modified-BPDN can much easily to apply its bound under very small $n$.
\item When $n$ is large enough, e.g. $n = 0.5m$ for $|N|=15=15\%m$, the bounds are almost equal for all values of $|\Delta|$ (the black plot of Fig. 2) including $|\Delta|=|N|$ (BPDN).
\end{enumerate}
\section{Conclusions}
We proposed a modification of the BPDN idea, called modified-BPDN, for sparse reconstruction from noisy measurements when a part of the support is known, and bounded its reconstruction error. A key feature of our work is that the bounds that we obtain are computable. Hence we are able to use Monte Carlo to show that the average value of the bound increases as the unknown support size or the size of the error in the known support increases. We are also able to compare with the BPDN bound and show that (a) for practical support sizes (equal to 10\% of signal size it holds under very strong assumptions (require more than 95\% random Gaussian measurements for the bound to hold) and (b) for smaller support sizes (e.g. 1.5\% of signal size), the BPDN bound is much larger than the modified-BPDN bound.
\section{appendix}
\subsection{Proof of Lemma 1}
Suppose $\text{supp(b)}\subseteq N_e$. We know the vectors $y-A c=y-A_{N_e}c_{N_e}$ and $A c-A b=A_{N_e}(b_{N_e}-c_{N_e})$ are orthogonal because $A_{N_e}'(y-A_{N_e}c_{N_e})=0$ using (\ref{lssolution}). Thus we minimize function $L(b)$ over all vectors supported on set $N_e$ by minimizing:
\begin{equation}
F(b)=\frac{1}{2}\|A_{N_e} c_{N_e}-A_{N_e} b_{N_e}\|_2^2+\gamma \|b_{T^c}\|_1
\end{equation}
Since this function is strictly convex, then $0\in \partial F(\tilde{b})$. Hence,
\begin{equation}
A'_{N_e}A_{N_e}\tilde{b}_{N_e}-A'_{N_e}A_{N_e} c_{N_e}+\gamma g_{N_e}=0
\end{equation}
Then, we have
\begin{equation}
c_{N_e}-\tilde{b}_{N_e}=\gamma(A'_{N_e}A_{N_e})^{-1}g_{N_e}
\end{equation}
Since
\begin{displaymath}
A'_{N_e}A_{N_e}= \left [ \begin{array}{cc}
A'_TA_T & A'_TA_{\Delta}\\
A'_{\Delta}A_T & A'_{\Delta}A_{\Delta}
\end{array} \right ]
\end{displaymath}
By using the block matrix inversion and $g_T=0$, we get
\begin{displaymath}
c_{N_e}-\tilde{b}_{N_e}= \left [ \begin{array}{c}
-\gamma (A'_TA_T)^{-1}A'_TA_{\Delta}(A'_{\Delta}MA_{\Delta})^{-1}g_{\Delta}\\
\gamma (A'_{\Delta}MA_{\Delta})^{-1}g_{\Delta})
\end{array} \right ]
\end{displaymath}
Thus, we can obtain the $l_{\infty}$ norm bound of error as below:
\begin{align}
&\|\tilde{b}_{N_e}-c_{N_e}\|_{\infty}=\gamma\|(A'_{N_e}A_{N_e})^{-1}g_{N_e}\|_{\infty} \nonumber\\
\le &\gamma \max( \|(A'_TA_T)^{-1}A'_TA_{\Delta}(A'_{\Delta}MA_{\Delta})^{-1}g_{\Delta}\|_{\infty},\nonumber\\
&\quad \quad \quad \quad \quad \|(A'_{\Delta}MA_{\Delta})^{-1}g_{\Delta}\|_{\infty}) \nonumber \\
\le &\gamma
\max(\|(A'_TA_T)^{-1}A'_TA_{\Delta}(A'_{\Delta}MA_{\Delta})^{-1}\|_{\infty}
,\nonumber\\
&\quad \quad \quad \quad \quad \|(A'_{\Delta}MA_{\Delta})^{-1}\|_{\infty})\nonumber
\end{align}
This follows using $\|g_{\Delta}\|_{\infty}=1$. Also, using $\|g_{\Delta}\|_2\le \sqrt{|\Delta|}$, we get the $l_2$ norm bound of $\tilde{b}-c$.

Using  $\|(A_T'A_T)^{-1}\|_2\le \frac{1}{1-\delta_{|T|}}$, $\|A_{\Delta}'A_{\Delta}\|_2\ge 1-\delta_{|\Delta|}$ and $\|A_T'A_{\Delta}\|_2\le \theta_{|T|,|\Delta|}$, we get (\ref{ripbound}).
\subsection{Proof of Lemma 2}
Suppose that $A_{N_e}$ has full column rank,
and let $\tilde{b}$ minimize the function $L(b)$ over all $b$ supported on $N_e=T\cup \Delta$. We need to prove under this condition,
$\tilde{b}$ is the unique global minimizer of $L(b)$.

The idea is to prove under the given condition, any small perturbation $h$ on
$\tilde{b}$ will increase function $L(\tilde{b})$,i.e.
$L(\tilde{b}+h)-L(\tilde{b})>0,\forall ||h||_{\infty} \le \delta$ for $\delta$ small enough. Since $L(b)$ is a convex function, $\tilde{b}$ should be the unique global minimizer.

Similar to \cite{justrelax}, we first split the perturbation into two parts $h=u+v$ where
$supp(u)=N_e$ and $supp(v)=N_e^c$. Clearly $||u||_{\infty}\le ||h||_{\infty}\le \delta$. Then we have
\begin{equation}
L(\tilde{b}+h)=\frac{1}{2}||y-A(\tilde{b}+u)-Av||_2^2+\gamma||(\tilde{b}+u)_{T^c}+v_{T^c}||_1 \label{disturbfunction}
\end{equation}
Then expand the first term, we can obtain
\begin{eqnarray}
\|y-A(\tilde{b}+u)-Av\|_2^2=\|y-A(\tilde{b}+u)\|_2^2+\|Av\|_2^2\nonumber\\
-2Re\langle y-A\tilde{b},Av \rangle +2Re \langle Au,Av \rangle
\end{eqnarray}
The second term of (\ref{disturbfunction}) becomes
\begin{equation}
\|(\tilde{b}+u)_{T^c}+v_{T^c}\|_1=\|(\tilde{b}+u)_{T^c}\|_1+\|v_{T^c}\|_1
\end{equation}
Then we have
\begin{eqnarray}
L(\tilde{b}+h)-L(\tilde{b})=L(\tilde{b}+u)-L(\tilde{b})+\frac{1}{2}\|Av\|_2^2\nonumber\\
-Re\langle y-A\tilde{b},Av \rangle +Re \langle Au,Av \rangle +\gamma\|v_{T^c}\|_1
\end{eqnarray}
Since $\tilde{b}$ minimizes $L(b)$ over all vectors supported on $N_e$, $L(\tilde{b}+u)-L(\tilde{b})\ge 0$. Then since $L(\tilde{b}+u)-L(\tilde{b})\ge 0$ and $\|Av\|_2^2 \ge 0$, we
need to prove that the rest are
non-negative:$\gamma\|v_{T^c}\|_1-Re \langle y-A\tilde{b},Av\rangle +Re\langle Au,Av \rangle \ge
0$. Instead, we can prove this by proving a stronger one
$\gamma\|v_{T^c}\|_1-|\langle y-A\tilde{b},Av\rangle |-|\langle Au,Av \rangle |\ge 0$.\\
Since $\langle y-A\tilde{b},Av\rangle =v'A'(y-A\tilde{b})$ and $supp(v)=N_e^c$,
\begin{equation}
|\langle y-A\tilde{b},Av\rangle |=|v_{N_e^c}'A_{N_e^c}'(y-A\tilde{b})|\le \|v\|_1\|A_{N_e^c}(y-A\tilde{b})\|_{\infty}\nonumber
\end{equation}
Thus,
\begin{eqnarray}
|\langle y-A\tilde{b},Av\rangle |
\le \max_{\omega \notin N_e}|\langle y-A\tilde{b},A_{\omega}\rangle |||v||_1
\end{eqnarray}
The third term of (\ref{disturbfunction}) can be written as
\begin{equation}
|\langle Au,Av\rangle |\le \|A'Au\|_{\infty}||v||_1\le
\delta\|A'A\|_{\infty}||v||_1
\end{equation}
And $\|v\|_1=\|v_{T^c}\|_1$ since $supp(v)=N_e^c \subseteq T^c$. Therefore,
\begin{equation}
L(\tilde{b}+h)-L(\tilde{b})\ge \big{[}\gamma-\max_{\omega \notin
N_e}|\langle y-A\tilde{b},A_{\omega}\rangle |-\delta||A'A||_{\infty}\big{]}||v||_1
\end{equation}
Since we can select $\delta>0$ as small as possible, then we just need to have
\begin{equation}
\gamma-\max_{\omega \notin N_e}|\langle y-A\tilde{b},A_{\omega}\rangle |>0
\label{directcond}
\end{equation}
Invoke Lemma 1, we have $A_{N_e}(c_{N_e}-\tilde{b}_{N_e})=\gamma
MA_{\Delta}(A'_{\Delta}MA_{\Delta})^{-1}g_{\Delta}$. Since
$y-A\tilde{b}=(y-A_{N_e}c_{N_e})+A_{N_e}(c_{N_e}-\tilde{b}_{N_e})$, therefore,
\begin{eqnarray}
|\langle y-A\tilde{b},A_{\omega}\rangle |\le |\langle y-A_{N_e}c_{N_e},A_{\omega}\rangle |\quad \quad \nonumber\\
+\gamma
|\langle (A'_{\Delta}MA_{\Delta})^{-1}A'_{\Delta}MA_{\omega},g_{\Delta}\rangle |
\end{eqnarray}
Then we only need to have the condition
\begin{eqnarray}
\gamma-\max_{\omega \notin N_e}
\big{[}\gamma|\langle (A'_{\Delta}MA_{\Delta})^{-1}A'_{\Delta}MA_{\omega},g_{\Delta}\rangle |+\nonumber\\
|\langle y-A_{N_e}c_{N_e},A_{\omega}\rangle |\big{]}>0\label{tightcond}
\end{eqnarray}
Since $y-A_{N_e}c_{N_e}$ is orthogonal to $A_w$ for each $\omega \in N_e$,
then $\max_{\omega \notin
N_e}|\langle y-A_{N_e}c_{N_e},A_{\omega}\rangle |=\|A'(y-A_{N_e}c_{N_e})\|_{\infty}$. Also, we know that $\max_{\omega \notin
N_e}|\langle (A'_{\Delta}MA_{\Delta})^{-1}A'_{\Delta}MA_{\omega},g_{\Delta}\rangle |\le \max_{\omega \notin
N_e}\|(A'_{\Delta}MA_{\Delta})^{-1}A'_{\Delta}MA_{\omega}\|_1\big{]}$. Thus, (\ref{tightcond}) holds if the following condition holds
\begin{equation}
||A'(y-A_{N_e}c_{N_e})||_{\infty}<\gamma \big{[}1-\max_{\omega \notin
N_e}||(A'_{\Delta}MA_{\Delta})^{-1}A'_{\Delta}MA_{\omega}
||_1\big{]}\label{directcond2}
\end{equation}
i.e. $\tilde{b}$ is the unique global minimizer if (\ref{directcond2}) holds.


\begin{thebibliography}{99}
\small{
\bibitem{BPDN} S. S. Chen, D. L. Donoho, and M. A. Saunders,\textit{\textbf{Atomic decomposition
by basis pursuit}}, SIAM J. Sci. Comput., vol. 20, no. 1, pp. 33-61, 1999.
\bibitem{justrelax} Joel A. Tropp, \textit{\textbf{Just Relax: Convex Programming Methods for Identifying Sparse Signals in Noise}}, IEEE Trans. on Information Theory, 52(3), pp. 1030 - 1051, March 2006.
\bibitem{modcs} Namrata Vaswani and Wei Lu, \textit{\textbf{Modified-CS: Modifying Compressive Sensing for Problems with Partially Known Support}}, IEEE Intl. Symp. Info. Theory (ISIT), 2009
\bibitem{modcsMRI} Wei Lu and Namrata Vaswani,\textit{\textbf{Modified Compressive Sensing for Real-time Dynamic MR Imaging}}, IEEE Intl. Conf. Image Proc (ICIP), 2009
\bibitem{analyzelscs} Namrata Vaswani,\textit{\textbf{Analyzing Least Squares and Kalman Filtered Compressed Sensing}}, IEEE Intl. Conf. Acous. Speech. Sig. Proc. (ICASSP), 2009.
\bibitem{decodinglp} E. Candes and T. Tao. \textit{\textbf{Decoding by Linear Programming}}, IEEE
Trans. Info. Th., 51(12):4203 - 4215, Dec. 2005.
\bibitem{iBPDN} L. Jacques, \textit{\textbf{A short Note on Compressed Sensing with Partially Known Signal Support}}, Arxiv preprint arXiv:0908.0660v1, 2009.
\bibitem{timevarying} D. Angelosante, E. Grossi, G. B. Giannakis,\textit{\textbf{Compressed Sensing of time-varying signals}}, DSP 2009
\bibitem{weightedl1} A. Khajehnejad, W. Xu, A. Avestimehr, B. Hassibi, \textit{\textbf{Weighted l1 Minimization for Sparse Recovery with Prior Information}}, IEEE Intl. Symp. Info. Theory(ISIT),2009
}
\end{thebibliography}
\end{document}